\documentclass[manuscript]{emulateapj}

\shorttitle{Oscillation of newly formed loops} \shortauthors{Yang \& Xiang}

\journalinfo{Accepted for publication in ApJL}

\submitted{Received 2016 January 24; accepted 2016 February 19}

\begin{document}

\title{OSCILLATION OF NEWLY FORMED LOOPS AFTER MAGNETIC RECONNECTION\\ IN THE SOLAR CHROMOSPHERE}

\author{Shuhong Yang\altaffilmark{1,2} and Yongyuan Xiang\altaffilmark{3}}

\altaffiltext{1}{Key Laboratory of Solar Activity, National
Astronomical Observatories, Chinese Academy of Sciences, Beijing
100012, China; shuhongyang@nao.cas.cn}
\altaffiltext{2}{College of
Astronomy and Space Sciences, University of Chinese Academy of
Sciences, Beijing 100049, China}
\altaffiltext{3}{Fuxian Solar Observatory, Yunnan Observatories,
Chinese Academy of Sciences, Kunming 650011, China}

\begin{abstract}

With the high spatial and temporal resolution H$\alpha$ images from the New Vacuum Solar Telescope, we focus on two groups of loops with a X-shaped configuration in the dynamic chromosphere. We find that the anti-directed loops approach each other and reconnect continually. The connectivity of the loops is changed and new loops are formed and stack together. The stacked loops are sharply bent, implying that they are greatly impacted by the magnetic tension force. When another more reconnection process takes place, one new loop is formed and stacks with the previously formed ones. Meanwhile, the stacked loops retract suddenly and move toward the balance position, performing an overshoot movement, which led to an oscillation with an average period of about 45 s. The oscillation of newly formed loops after magnetic reconnection in the chromosphere is observed for the first time. We suggest that the stability of the stacked loops is destroyed due to the join of the last new loop and then suddenly retract under the effect of magnetic tension. Because of the retraction, another lower loop is pushed outward and performs an oscillation with the period of about 25 s. The different oscillation periods may be due to their difference in three parameters, i.e., loop length, plasma density, and magnetic field strength.

\end{abstract}

\keywords{magnetic reconnection --- Sun: chromosphere --- Sun:
evolution --- Sun: oscillations}

\section{INTRODUCTION}

Magnetic reconnection is widely considered to play an important role in driving many types of astrophysical events, such as solar and stellar
flares, and accretion disks (Haisch et al. 1991; Shibata et al. 1995; Yokoyama \& Shibata 1995; Yuan et al. 2009). Due to magnetic reconnection, magnetic topology is changed, i.e., the loops involved in the reconnection disappear and new loops are formed (Zweibel \& Yamada 2009). When magnetic reconnection takes place, free energy is released and converted to kinetic and thermal energies of plasma. In the solar corona, many signatures of magnetic reconnection have been widely observed, e.g., reconnection inflows and outflows (Innes et al. 2003; Asai et al. 2004; Takasao et al. 2012), formation of post flare loops (Su et al. 2013; Sun et al. 2015), cusp-shaped structures above flare loops (Tsuneta et al. 1992; Masuda et al. 1994), and current sheets (Lin et al. 2005). In addition, experiments which are dedicated to magnetic reconnection have been carried out in laboratories under controlled conditions, and magnetic reconnection process similar to that in a solar flare was simulated successfully (Bratenahl \& Yeates 1970; Zhong et al. 2010). Besides the large-scale solar activities in the corona, small-scale reconnection in the solar chromosphere was also evidently observed with the H$\alpha$ observations (Yang et al. 2015). In their study, the results reveal that slow reconnection triggered the instability around the reconnection site and led to a rapid reconnection.

In previous observations, newly formed loops in the magnetic reconnection retracted from the reconnection site to a new position, e.g., retraction of post flare loops to a lower height in the corona, or separation of two groups of newly formed loops in the chromosphere.
White et al. (2012) observed a transverse loop oscillation in the hot coronal lines following a coronal mass ejection. They suggested that magnetic
reconnection may create a post flare hot loop and then release the loop downwards the solar surface, thus leading to the transverse oscillation.
In the present paper, we will report an oscillation behavior of newly formed loops after magnetic reconnection in the solar chromosphere observed with the New Vacuum Solar Telescope (NVST; Liu et al. 2014). The NVST is a 1 meter aperture vacuum telescope in \emph{Fuxian Solar Observatory} of China. The H$\alpha$ line is adopted by NVST to image the small-scale structures in the chromosphere at high spatial and temporal resolutions, which is helpful for us to investigate the highly dynamic structures therein.

\section{OBSERVATIONS AND DATA ANALYSIS}

\begin{figure}
\centering
\includegraphics
[clip,angle=0,width=0.48\textwidth]{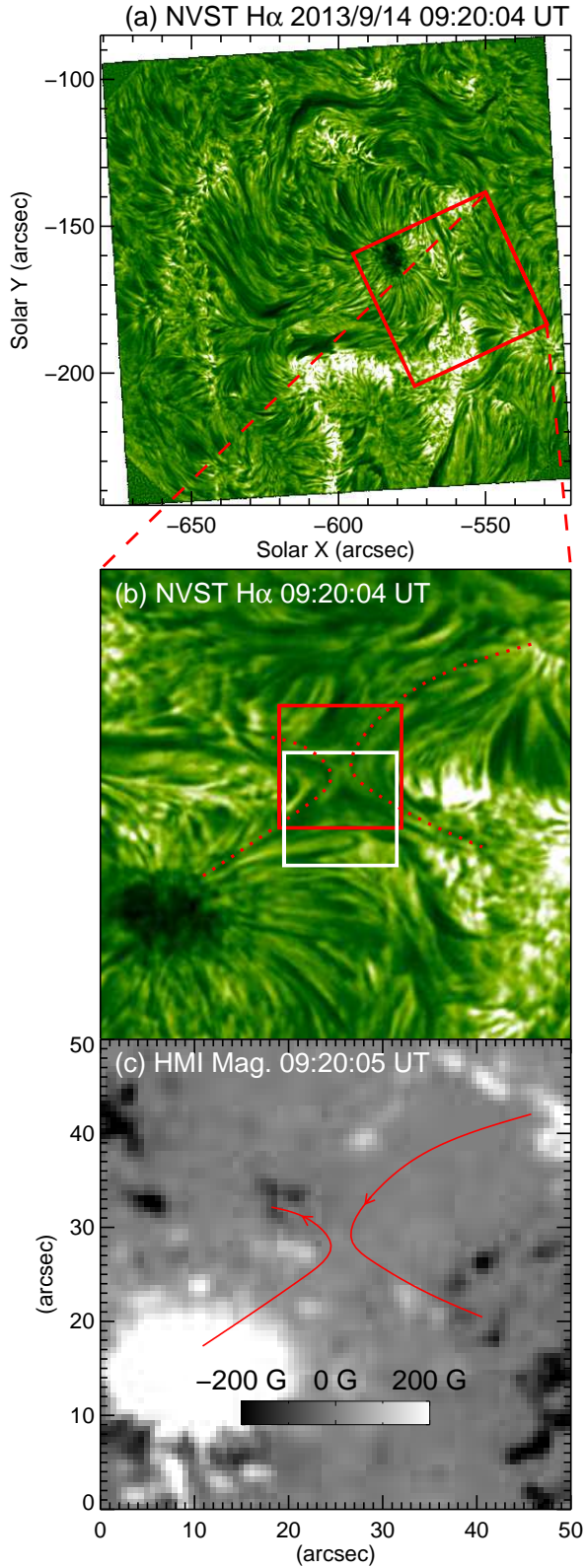} \caption{(a)
Overview of NOAA AR 11841 observed with the NVST in H$\alpha$ 6562.8
{\AA} line on 2013 September 14. (b and c) expanded H$\alpha$ image
and the corresponding photospheric magnetogram in the FOV outlined by
the square in panel (a). The red window in panel (b) outlines the
FOV of Figures 2 and 3, and the white window delineates the FOV of
Figure 4. The red curves outline two groups of loops involved in
reconnection. \label{fig1}}
\end{figure}

The NVST data used in this study were obtained in H$\alpha$ 6562.8 {\AA} line on 2013 September 14. The H$\alpha$ images have a pixel size of 0$\arcsec$.163 and a field-of-view (FOV) of 151$''$ $\times$ 151$''$, covering NOAA AR 11841 and its nearby region (see Figure 1(a)). The H$\alpha$ images were obtained from 06:35 UT to 09:35 UT at a rate of 8.3 frames per second. All the Level 0 data were flat field corrected and dark current subtracted first, and then every one hundred Level 1 frames were reconstructed to one Level 1+ image by speckle masking (Weigelt 1977; Lohmann et al. 1983). The reconstructed images have a cadence of 12 s. In order to study the oscillation process of the newly formed loop, we adopt the H$\alpha$ images with a cadence of 3 s between 09:19:58 UT and 09:26:58 UT.

In addition, we also use the simultaneous Atmospheric Imaging Assembly (AIA; Lemen et al. 2012) extreme-ultraviolet images and the Helioseismic and Magnetic Imager (HMI; Scherrer et al. 2012) magnetograms from the \emph{Solar Dynamics Observatory} (\emph{SDO}; Pesnell et al. 2012). The cadence and pixel size of the AIA images are 12 s and 0$\arcsec$.6, while those of the HMI magnetograms are 45 s and 0$\arcsec$.5, respectively. We calibrated the AIA and HMI data to Level 1.5 with the standard procedure \emph{aia\_prep.pro}, and rotated them differentially to a reference time (09:22:00 UT). Then, the NVST H$\alpha$ images were co-aligned to \emph{SDO} data by the cross-correlation method.

\section{RESULTS}

\begin{figure}
\centering
\includegraphics
[bb=133 256 443 563,clip,angle=0,width=0.47\textwidth]{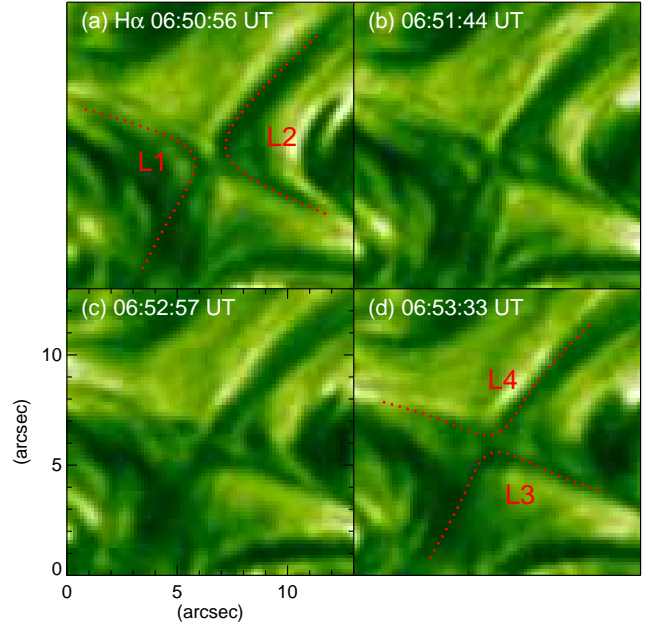}
\caption{Sequence of H$\alpha$ images showing a reconnection process
(also see the accompanying animation Movie 1). Curves ``L1" and ``L2" outline the
initial loops before reconnection, and curves ``L3" and
``L4" denote the newly formed loops after reconnection.
\label{fig1}}
\end{figure}

\begin{figure*}
\centering
\includegraphics
[bb=93 259 489
580,clip,angle=0,width=0.85\textwidth]{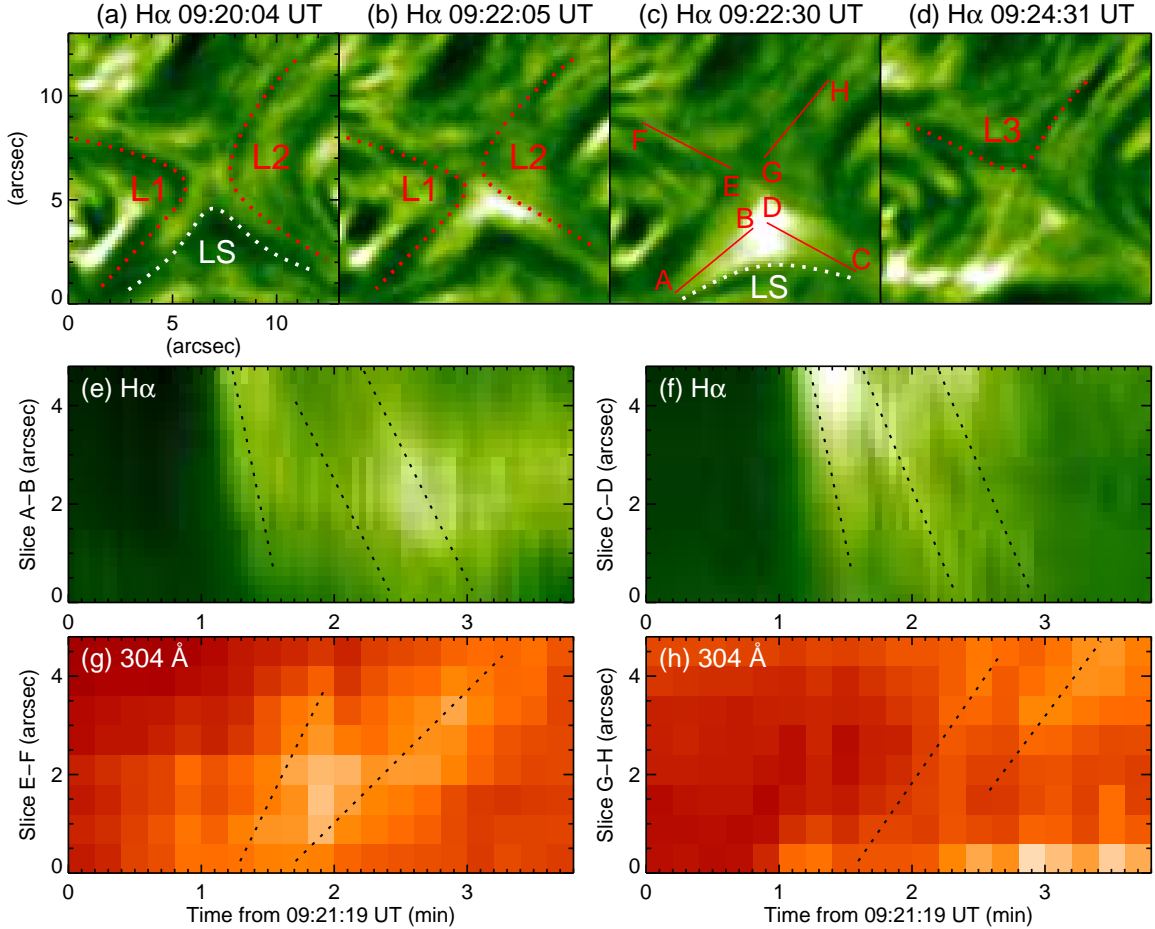}
\caption{(a-d) Sequence of H$\alpha$ images showing another
reconnection process (also see the accompanying animation Movie 2).
(e-f) Space-time plots derived from H$\alpha$ images along slices
``A-B", and ``C-D" (marked in panel (c)), respectively. (g-h)
Space-time plots derived from AIA 304 {\AA} images along slices
``E-F", and ``G-H", respectively. Curves ``L1" and ``L2" delineate
the flux loops before reconnection, and curves ``L3" and ``LS"
outline the newly formed loops. The dotted lines in panels (e)-(h)
follow the bright moving features. \label{fig1}}
\end{figure*}

\begin{figure*}
\centering
\includegraphics
[bb=90 134 491
686,clip,angle=0,width=0.72\textwidth]{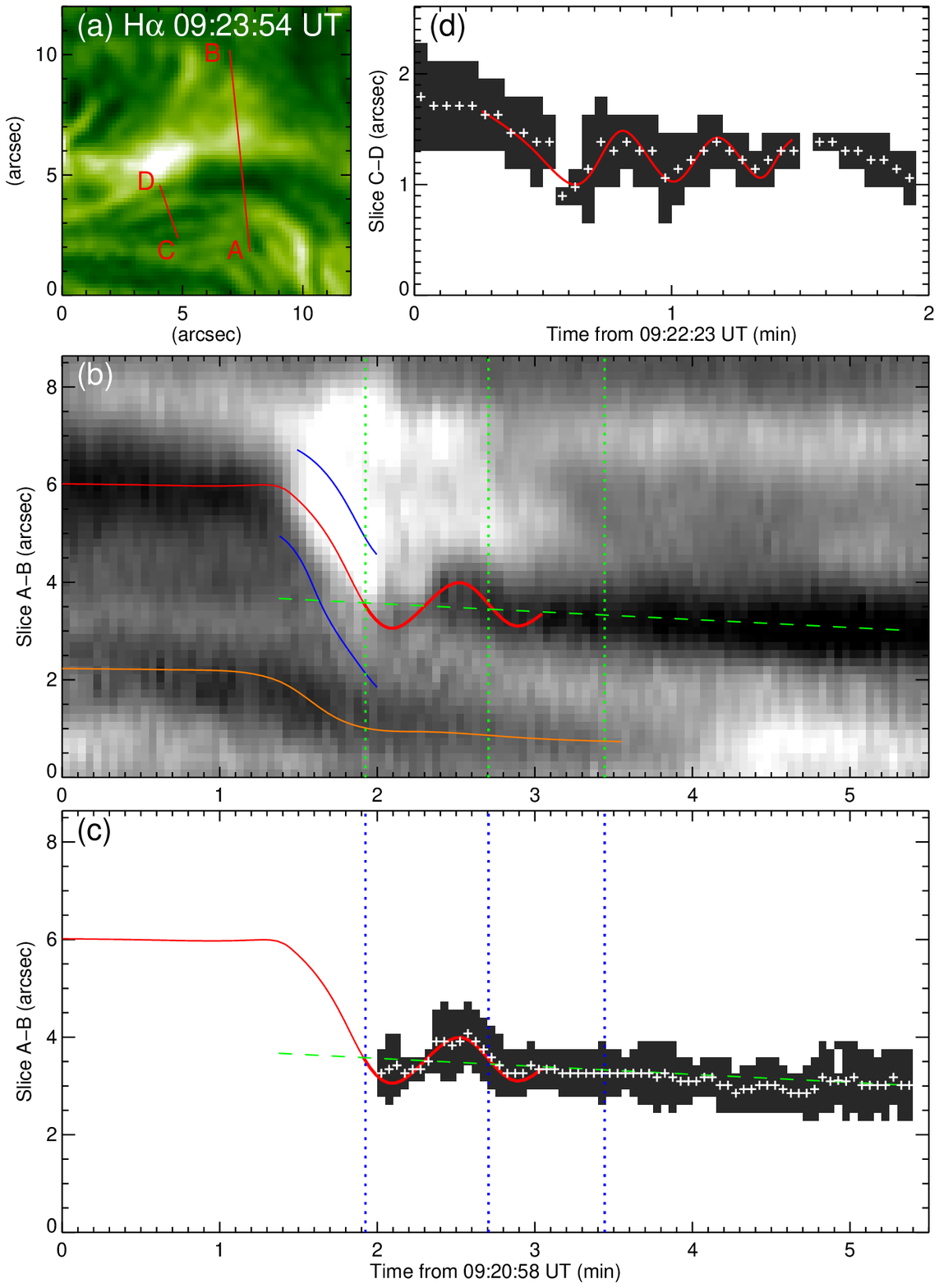}
\caption{(a) H$\alpha$ image observed at 09:23:54 UT. (b-c)
Space-time plot derived from H$\alpha$ images along slice ``A-B"
(marked in panel (a)), and corresponding filtered image
by an unsharp-mask sharpening filter showing the oscillation of the loop set.
The blue and red curves delineate the two edges and the main body of the
oscillating loop set. The dashed line marks the balance position
of the oscillation, and the vertical dotted lines separate two
oscillation periods. The brown curve outlines the trajectory of a
lower loop. (d) Unsharp-masked image of the space-time
plot derived from H$\alpha$ images along slice ``C-D".
The red curve in panel (d) delineates the main body of
another oscillating loop. Each ``+" symbol in panels (c)
and (d) marks the center of the dark feature in each time slice.
An animation (Movie 3) of this figure is available.
\label{fig1}}
\end{figure*}

The region of interest is located at the edge of the main sunspot of AR 11841 and is outlined by the window in Figure 1(a). The H$\alpha$ image is anticlockwise rotated 65$\degr$, the expanded view of which is shown in panel (b). In the H$\alpha$ image, two groups of dark fibrils (delineated by the red dotted curves) form a X-shaped configuration, which might reveal the magnetic topology according to the previous studies (e.g., Yang et al. 2014, 2015). The left loops originate from the main sunspot with positive polarity to the nearby negative fields (see panel (c)). The right loops connect the small-scale fields with opposite polarity, and have an anti-direction to the left ones.

At the X-point of the anti-directed loops, we observe some reconnection processes. In order to display the reconnection clearly, we focus on the small area outlined by the red window in Figure 1(b). One process of reconnection is presented in Figure 2. At 06:50:56 UT, two loops (outlined by dotted curves ``L1" and ``L2") had a distance of about 1{\arcsec}.5 (see Figure 2(a)). They approached each other, and 48 s later, their distance became much shorter, no more than 1{\arcsec} (panel (b)). Then they continued to move toward each other and interacted eventually (see panel (c)). Due to the reconnection (also see the accompanying animation Movie 1), the former loops ``L1" and ``L2" disappeared, and two new loops ``L3" and ``L4" were formed (denoted by the red curves in panel (d)).

At the same reconnection site, many reconnection processes occurred continually, and more and more loops were formed and stacked together (marked by the white curve ``LS" in Figure 3(a)). Figures 3(a)-(d) show another reconnection between loops ``L1" and ``L2" (also see the accompanying animation Movie 2). Before the reconnection, the distance between loops ``L1" and ``L2" was about 2{\arcsec} (see panel (a)). Then two minutes later, loops ``L1" and ``L2" got much closer (panel (b)). They went on approach and reconnected, and one new loop was formed and stacked with the previously formed ones, marked by curve ``LS" in panel (c). The loops ``LS" were released from near the reconnection site and moved outward, and performed a behavior of oscillation (see the lower part of the animation Movie 2). The dotted curve in panel (d) outlines the other new loop formed due to the reconnection. Along slices ``A-B" and ``C-D" marked in panel (c), two space-time plots are derived from the H$\alpha$ images and displayed in panels (e) and (f), respectively. In panels (e) and (f), there are many bright features moving from ``B" to ``A" and from ``D" to ``C", as indicated by the dotted lines. At the early stage of the reconnection, the mean velocity of the bright features is 158 km s$^{-1}$. The following bright features have an average velocity of about 75 km s$^{-1}$. Along slices ``E-F" and ``G-H", we make two space-time plots with AIA 304 {\AA} observations and show them in panels (g) and (h). In the 304 {\AA} plots, bright moving features can be identified from ``E" to ``F" and from ``G" to ``H" (marked by dotted lines). The mean velocity along ``E-F" and ``G-H" is about 50 km s$^{-1}$.

To show the oscillations of newly formed loops well, we use the H$\alpha$ images with the cadence as high as three minutes (also see the accompanying animation Movie 3). Along slice ``A-B" (as marked in Figure 4(a)), we make a space-time plot and display it in panel (b). The two blue curves delineate the two edges of the loop set, and the red curve outlines the center of the main body. The loops were initially located at the position of about 6{\arcsec} in slice ``A-B" at 09:20:58 UT. About one and a half minutes later, half of the loops which are close to the reconnection site got much brighter. Thereafter, the whole set of loops began to move from ``B" to ``A" with a mean velocity of 60 km s$^{-1}$. When they reached the balance position (dashed line), they did not stop and went on move toward ``A" until they reached the farthest position of about 0.{\arcsec}6 away from the balance position, appearing as an overshoot movement. Then the loops began to move back toward ``B", and then performed a convergent oscillation. We can identify one and a half oscillation periods, as separated by the vertical dotted lines. The average period is about 45 s. Due to the sudden retraction of the loops from near the reconnection site, one lower loop (see the brown curve) was pushed toward ``A", as revealed in the lower part of panel (b). To show the oscillation well, we apply the \emph{unsharp\_mask} procedure to the space-time plot in panel (b) under IDL, and display the filtered image in panel (c). Each ``+" symbol marks the center of the dark feature in each time slice. We can see that the filtered curve reveals a periodic oscillation, which agrees with the red curve. Along slice ``C-D" (marked in panel (a)), another space-time plot is derived from the H$\alpha$ data and the corresponding sharp-masked image is presented in panel (d). This loop was pushed and oscillated with the period of about 25 s (see the red curve in the top-right panel).

\section{CONCLUSIONS AND DISCUSSION}

With the high resolution H$\alpha$ images from the NVST, we focus on two groups of loops with a X-shaped configuration in the dynamic chromosphere. We find that the anti-directed loops approached each other and then reconnected continually. The magnetic topology was changed, and many new loops were formed and stacked together. At last, as one more reconnection process occurred, a new loop was formed and stacked with the loop set. Then this set of loops suddenly retracted toward the balance position and performed an overshoot movement, leading to a convergent oscillation with the mean period of about 45 s. Due to the sudden retraction, another lower loop was pushed outward and performed an oscillation with the period of about 25 s.

With a 2.5-dimensional MHD simulation method, Murray et al. (2009) found that oscillatory reconnection, a series of reconnection reversals, will take place under the conditions of quasi-bounded outflow regions during the reconnection bursts. The oscillatory reconnection is initiated in a self-consistent manner instead of driven manner. In a further study of McLaughlin et al. (2012), two distinct periodic phases in the oscillatory oscillation were identified for the first time, i.e., the transient, impulsive phase, and the stationary phase. During the observations of solar flares, very short period radio pulsations were found in many studies (e.g., Kliem et al. 2000; Tan et al. 2007; also the review by Nakariakov \& Melnikov 2009). The radio pulsations were interpreted to be caused by periodic particle acceleration episodes which resulted from a highly dynamic magnetic reconnection regime in an extended large-scale current sheet. Although oscillatory reconnection may be not directly linked with the loop set oscillation presented in our study, it may contribute to the continuous formation of new loops that stack together.

For the event in this paper, magnetic reconnection took place between two sets of anti-directed loops, the footpoints of which form a quadrupolar field (see Figure 1). Similar X-shaped geometries in the chromosphere have been observed by Yang et al. (2015) and Xue et al. (2016). All of these events were located at the edges of active regions. For each X-shaped structure, one or more footpoints were found to coincide with the strong sunspot fields, and the others were at the nearby small-scale fields. Slow magnetic reconnection can occur continually at the X-point. In Yang et al. (2015), the slow reconnection step was found to be at least several tens of minutes, and in the present paper, the slow reconnection lasted much longer. As revealed by numerical simulations, reconnection rate is relevant with many parameters of plasma, such as plasma $\beta$, and Lundquist number (Ni et al. 2012, 2013). Continuous slow reconnection can be sustained if plentiful loops with opposite directions approach each other successively.

In the present study, two groups of oppositely directed loops formed a X-shaped structure, a configuration where magnetic reconnection always takes place. Indeed we observe continuous reconnection between them, as shown in Figures 2 and 3 (also see Movies 1 and 2). Magnetic reconnection resulted in a change of magnetic connectivity, and outflows of heated plasmas (revealed by the outward motion of bright features from the reconnection site). As the reconnection took place, the newly formed loops stacked together (as denoted by the dotted curve ``LS" in Figure 3). The stack of new loops was also observed by Yang et al. (2015) (see Figure 2 therein). When another more new loop was formed and stacked with the former ones (see Figure 3 in the present paper), the stability was destroyed and thus the loop set retracted suddenly, similar to an ``avalanche." We note that, the stacked loops ``LS" in Figure 3(a) were sharply bent compared with those in Figure 3(c), implying that the loop set was greatly impacted by the magnetic tension force. In the studies of White et al. (2012, 2013), magnetic reconnection was considered to play an important role in the excitation of transverse loop oscillations. Our observations also indicate that the excitation mechanism for the loop set oscillation is linked to magnetic reconnection which form the new loops and release them outward, instead of being hit by an external blast wave (Nakariakov et al. 1999; Verwichte et al. 2010; Wang et al. 2012).

The retracted loops moved toward the balance position, and overshot to a farther site. Then under the effect of magnetic tension, they moved back and oscillated with the period of about 45 s. This kind of oscillation performance of newly formed loops after magnetic reconnection in the solar chromosphere has not been observed in the previous studies. Due to the retraction, one lower loop oscillated with the period of about 25 s.
The different oscillation periods mainly result from the difference in three parameters, i.e., loop length, plasma density, and magnetic field strength, according to the previous studies (Aschwanden et al. 1999; Nakariakov \& Ofman 2001).

\acknowledgments { We thank the referees for their constructive comments and helpful suggestions. This work is supported by the National Natural Science Foundations of China (11533008, 11203037, 11221063, 11373004, and 11303049), the Strategic Priority Research Program$-$The Emergence of Cosmological Structures of the Chinese Academy of Sciences (No. XDB09000000), the CAS Project KJCX2-EW-T07, the Young Researcher Grant of National Astronomical Observatories of CAS, and the Youth Innovation Promotion Association of CAS (2014043). The data are used courtesy of NVST, HMI, and AIA science teams.\\}

{}

\end{document}